\documentclass[a4paper,onecolumn,11pt,allowtoday,unpublished]{quantumarticle}
\pdfoutput=1
\usepackage[utf8]{inputenc}
\usepackage[english]{babel}
\usepackage[T1]{fontenc}
\usepackage{amsmath}
\usepackage{amsmath,amssymb,amsthm,mathtools}
\usepackage{microtype,enumitem}
\usepackage[hidelinks]{hyperref}
\usepackage{comment}
\newtheorem{theorem}{Theorem}[section]
\newtheorem{lemma}[theorem]{Lemma}

\theoremstyle{definition}
\newtheorem{definition}[theorem]{Definition}
\newcommand{\ket}[1]{\lvert#1\rangle}
\newcommand{\bra}[1]{\langle#1\rvert}
\newcommand{\norm}[1]{\lVert#1\rVert}

\title{\color{black}Unitary Synthesis with a Unary Tracker}\author{Fred Sun}
\email{f4sun@uwaterloo.ca}
\affiliation{Institute for Quantum Computing, University of Waterloo}
\affiliation{School of Computer Science, University of
Waterloo}
\date{\today}
\begin{document}
\newcommand{\anton}[1]{{\color{red}(anton): #1}}
\newcommand{\fred}[1]{{\color{blue}(fred): #1}}
\maketitle
\hypersetup{
    pdftitle={Unitary Synthesis with a Unary Tracker},
    pdfauthor={Fred Sun},
    pdfencoding=unicode
}
\begin{abstract}
This manuscript studies general unitary synthesis using unary-encoded ancillary register. Our first construction focuses on the $T$-cost of unitary synthesis. We maintain a unary register that coherently tracks the computational-basis label of the data. This allows Boolean lookups to use only CNOTs and shares the precision-dependent synthesis cost across the branches of each cosine-sine decomposition (CSD) factor. For a classically specified $n$-qubit unitary with $K$ mixing branches in its CSD, we obtain $T$-count $O(K+NL)$ using $N+O(L)$ clean ancillas, where $N=2^n$ and $L=n+\log(1/\epsilon)$. We also give a related construction that directly applies the unitary onto a unary register to implement disjoint branches in parallel. Combining this with the quantum Shannon decomposition gives space-depth volume $O(N^2)$ using arbitrary single-qubit gates and CNOTs, which matches the asymptotically optimal bound.
\end{abstract}

\section{Introduction}
\label{sec:introduction}

Unitary synthesis is the problem of constructing a circuit of elementary quantum gates that implements a classically specified unitary. Its circuit complexity measures the resources required to implement this transformation.

We work with the universal Clifford+$T$ gate set. In fault-tolerant quantum computing, $T$ gates can be substantially more expensive than Clifford gates because of the cost of magic-state distillation. The $T$-count is therefore an important resource metric, and reducing it can be worthwhile even when this requires additional ancillas or Clifford operations \cite{LKS24,GKW24}.

Fault-tolerant implementations also require physical qubits and repeated error correction throughout the computation. Their cost therefore depends on both the space occupied and the time for which it is needed, motivating \textit{space-depth volume}, defined as the total number of qubits multiplied by the total circuit depth, as a resource metric \cite{Litinski19}. 

With these resource metrics in mind, we study how the structure of a unitary affects its synthesis $T$-count. We also consider how to reduce the ancilla requirements of our construction while retaining its asymptotic space-depth volume.

\paragraph{Prior work.}
An important ingredient in unitary synthesis is loading classically specified information into a quantum circuit. Low, Kliuchnikov, and Schaeffer studied how additional workspace can reduce the $T$-cost of this lookup \cite{LKS24}. Gosset, Kothari, and Wu used boolean functions to select short single-qubit gate sequences, obtaining optimal $T$-counts for state preparation and diagonal unitary synthesis \cite{GKW24}. Tan applied this programming approach to general unitary synthesis using the cosine-sine decomposition (CSD), grouping bits together to reduce the cost of boolean function computation \cite{Tan25}. Other approaches use block flattening and dilation or block-sparse Hamiltonian simulation to improve the general upper bounds \cite{YZZ26,Liu26}.

Our construction follows the CSD and boolean function-based approach. The CSD writes an $n$-qubit unitary as a sequence of multiplexed single-qubit operations: each multiplexer acts on one target qubit, while the other bits select which single-qubit matrix is applied \cite{SMB06,Tan25}. We call each selected single-qubit rotation a \textit{branch}. There are $O(2^n)$ such factors, and Boolean lookups select the gate instructions for their branches. This makes the cost of these lookups an important part of the synthesis cost.

\paragraph{Our approach.}

The current lower bound is $\Omega(2^n)$ $T$ gates for general $n$-qubit unitary synthesis at constant approximation error \cite{GKW24}. This motivates us to ask when the CSD approach can reach this scale, up to precision factors. Since there are $O(2^n)$ factors, our strategy is to make their Boolean lookups free in $T$-count and account separately for the cost of maintaining the lookup addresses.

We keep a unary register that coherently tracks the computational-basis label of the data register. On a unary address, any classically specified Boolean function can be loaded using only CNOTs, so these lookups have zero $T$-cost. We show the entire factor can be synthesized using $O(\log(1/\delta))$ $T$ gates for approximation error $\delta$. The question is then how to maintain this address throughout the entire CSD.

We observe that within each factor of the CSD, only one data bit can change. We therefore only need to update the pairs of unary wires corresponding to the two values of that target bit. A diagonal branch leaves the label unchanged. A purely off-diagonal branch flips the target bit, so an ordinary SWAP updates its unary label. A mixing branch can create a superposition of the two target values, and we use a constant number of Fredkin gates per mixing branch to remove the old target information and restore the new information.

The resulting $T$-count has two contributions: implementing the programmed operations and maintaining the unary tracker. The first is paid once per CSD factor, while the second depends on the number of mixing branches. This brings the bound towards the $2^n$ scale, up to precision factors, when the supplied factorization has sufficiently few mixing branches.

\paragraph{Main result.}
Write $N=2^n$, $L=n+\log{\frac{1}{\epsilon}}$ and let $K$ denote the number of mixing branches for the CSD of some unitary $U \in U(N)$. We obtain the following bound.

\begin{theorem}[Mixing-sensitive unitary synthesis]
\label{thm:main}
Let $n\ge1$ and $0<\epsilon<1/2$. Given a classically specified $U\in U(N)$ with $K$ mixing branches in its CSD, there is a measurement-free Clifford+$T$ circuit $C$ using $a=N+O(L)$ ancillas and $T$-count
\begin{align}
&T(C)=O\!\left(K+NL\right). \label{eq:intro-main-bound}
\end{align}
For every normalized input $\ket{\psi}$, it satisfies
\begin{align}
&\norm{C(\ket{\psi}\ket{0^a})-(U\ket{\psi})\ket{0^a}}\le\epsilon. \label{eq:intro-error}
\end{align}
We assume access to an ideal phase-gradient catalytic state.
\end{theorem}

The count $K$ depends on the supplied factorization, rather than being an invariant of the unitary. When $K$ is small, the tracker updates are inexpensive. However, for a dense factorization, $K$ can be $\Theta(N^2)$, so the bound does not give a general $O(2^n)$ synthesis algorithm, nor a general asymptotic improvement over SOTA.

Compared with synthesizing each CSD branch separately, our construction shares the precision-dependent cost across all branches of a factor. This replaces the naive bound $O(N^2L)$ with $O(K+NL)$. Even for the worst-case $K\in\Theta(N^2)$, the precision factor therefore multiplies $N$ rather than $N^2$. 

This reduction in $T$-count comes at the cost of $O(N)$ ancillas, and the CNOT lookups can still require substantial circuit depth. We therefore consider a variant that uses a smaller reusable unary register and study how its depth changes with the available workspace.

\paragraph{Space-depth volume.}
In this variant, we transfer the quantum state itself into unary form, rather than keeping a unary tracker alongside the binary data. Operations on disjoint pairs of unary wires can then run in parallel.

Encoding the entire state in unary form still requires $N$ wires. To reduce this requirement, we instead decompose the unitary into smaller blocks and use unary encoding only within each block. We use the quantum Shannon decomposition (QSD), which builds on the CSD to express a unitary in terms of smaller uncontrolled unitaries and multiplexed single-qubit rotations \cite{SMB06}. We stop the recursion when the blocks fit within the available workspace.

Each block is encoded, implemented, and decoded back into binary form, leaving the unary register empty for the next block. A smaller register therefore saves ancillas but requires more blocks to be executed sequentially. Counting these steps together with the rotations between blocks gives the following tradeoff.

\begin{theorem}[Space-depth tradeoff]
\label{thm:volume}
Let $N=2^n$ and $n\le a\le N$. Our construction uses at most $a$ ancillas to implement any $n$-qubit unitary. With arbitrary single-qubit gates and CNOTs, and assuming unrestricted connectivity, the implementation is exact and has depth and space-depth volume
\begin{align*}
D=O(N^2/a), \qquad V=O(N^2),
\end{align*}
respectively.
\end{theorem}
This variant attains the known asymptotically optimal space-depth volume over this ancilla range \cite{Sun23,Yuan26}.

The remainder of this manuscript proceeds as follows. In Sec.~\ref{sec:prelim}, we introduce the notation and the lookup and synthesis ingredients. These building blocks are combined to prove Theorem~\ref{thm:main} in Sec.~\ref{sec:main_sec}. Theorem~\ref{thm:volume} is then proved in Sec.~\ref{sec:volume}. We provide a few open questions in Sec.~\ref{sec:conclusion} for future work.

\section{Preliminaries}\label{sec:prelim}

Throughout, let $U\in U(2^n)$ be classically specified, and write
$N=2^n$. The data register $\mathsf D$ contains an arbitrary
state
\begin{align}
\ket{\psi}=\sum_{x\in\{0,1\}^n}\alpha_x\ket{x}.
\end{align}

\subsection{Unary tracking and classical lookup}

\begin{definition}[Unary tracker]
\label{def:tracker}
For $x\in\{0,1\}^n$, let $e_x\in\{0,1\}^N$ be a register of Hamming weight $1$, with the excitation in bit position $x$. An $N$-qubit register $\mathsf A$ coherently
tracks the data when their joint state has the form
\begin{align}
\sum_x\alpha_x\ket{x}_{\mathsf D}\ket{e_x}_{\mathsf A}.
\label{eq:tracked-state}
\end{align}
\end{definition}

The motivation for keeping this register is that it makes classical lookup free in the $T$-count model, as will be described in Lemma~\ref{lem:lookup}.

\begin{lemma}[Free boolean functions]
\label{lem:lookup}
For any classically specified boolean function
$f:\{0,1\}^n\to\{0,1\}^w$, the map
\begin{align*}
\ket{e_x}\ket{y}\longmapsto\ket{e_x}\ket{y\oplus f(x)}
\end{align*}
on inputs of Hamming weight $1$ can be implemented using only CNOTs. The output can be erased by reversing the same gates if the unary address is unchanged.
\end{lemma}

\begin{proof}
For each output bit, apply a CNOT from unary wire $x$
exactly when the corresponding bit of $f(x)$ is $1$.
Only one unary wire is occupied, so the accumulated XOR is
precisely $f(x)$. The address is untouched, and the same
CNOTs reverse the map.
\end{proof}

We use this lemma to load two tables: the gate instructions for
a single-qubit $SU(2)$ rotation, and a binary approximation of a phase.
The tables are computed classically from $U$. Their lookups
may use many CNOTs, but no $T$ gates.

\subsection{The CSD and its branches}

\begin{definition}[Multiplexed single-qubit unitary]
A unitary with target qubit $t$ is multiplexed if
\begin{align*}
V=\sum_{z\in\{0,1\}^{n-1}}
\ket{z}\bra{z}_{\bar t}\otimes V_z,
\qquad V_z\in U(2),
\end{align*}
where $\bar t$ denotes the other data qubits.
We call $V_z$ a \textit{branch} and $z$ its \textit{control string}.
\end{definition}

The cosine-sine decomposition (CSD) gives a factorization of any arbitrary unitary $U\in U(2^n)$ as follows:
\begin{align}
U=V_{N-1}\cdots V_1,
\label{eq:csd}
\end{align}
where each factor is a multiplexed single-qubit unitary \cite{SMB06}. 

For iteration $j$ of the CSD, we write the multiplexor as
\begin{align}
    V_j = \sum_{z\in\{0,1\}^{n-1}} \ket{z}\bra{z}_{\bar t_j} \otimes V_{j,z}.
\end{align}
The target qubit is the $t_j$th least significant bit in $D$, the controls of the multiplexor acts on all bits but $t_j$, namely $\bar t_j$, and the branches are labeled as $V_{j,z}$. Within an iteration $j$, we rewrite our data register as $\ket{z_j,s_j}$, with control string $z_j$ and target bit $s_j$. 
The corresponding pair of unary wires are $\ket{q_{z_j,0}}$ and $\ket{q_{z_j,1}}$.

\begin{definition}[Mixing branch count]
A branch is \emph{mixing} if its matrix has both nonzero diagonal
and nonzero off-diagonal entries. Let $K_{j,z} \in \{0,1\}$ be a binary indicator that equals $1$ if $V_{j,z}$ is a mixing branch. The total number of mixing branches is therefore
\begin{align}
K=\sum_{j=1}^{N-1}\sum_{z\in\{0,1\}^{n-1}}K_{j,z}.
\end{align}
This count refers to the supplied factorization
\eqref{eq:csd}.
\end{definition}

We label each unitary $U\in U(2)$ \textit{diagonal}, \textit{off-diagonal}, or \textit{mixing}.
Diagonal branches change phases without changing the target bit.
Off-diagonal branches flip the bit and may also change phases.
Mixing branches can create a superposition of both target values.
The following factorization lets us handle these cases separately.

\begin{lemma}[Separating target operations from phases]
\label{lem:factor}
Each branch can be written as a target operation followed by
a phase correction:
\begin{equation}
V_{j,z}=D_{j,z}R_{j,z},
\label{eq:branch-factor}
\end{equation}
where \(D_{j,z}\in U(2)\) is diagonal. For a diagonal,
off-diagonal, or mixing branch, we take \(R_{j,z}\) to be
\(I\), \(X\), or an element of \(SU(2)\), respectively.
In the mixing case, the phase correction is just
\(D_{j,z}=e^{i\phi_{j,z}}I\) for some real \(\phi_{j,z}\).
\end{lemma}

\begin{proof}
For a diagonal branch, no target operation is needed:
take \(R_{j,z}=I\) and \(D_{j,z}=V_{j,z}\).

For an off-diagonal branch, multiplying by \(X\) makes the
matrix diagonal. Thus take \(R_{j,z}=X\) and
\(D_{j,z}=V_{j,z}X\). Since \(X^2=I\), we have
\(D_{j,z}R_{j,z}=V_{j,z}\).

For a mixing branch, we remove a single phase to make the
determinant one. Since \(V_{j,z}\) is unitary, write
\(\det V_{j,z}=e^{2i\phi_{j,z}}\) and set
\[
R_{j,z}=e^{-i\phi_{j,z}}V_{j,z},\qquad
D_{j,z}=e^{i\phi_{j,z}}I.
\]
Then \(D_{j,z}R_{j,z}=V_{j,z}\), and
\[
\det R_{j,z}
=e^{-2i\phi_{j,z}}\cdot \det V_{j,z}=1.
\]
Hence \(R_{j,z}\in SU(2)\).
\end{proof}
For each factor, write $V_j=D_jR_j$, where
\begin{align}
R_j&=\sum_z\ket{z}\bra{z}_{\bar t_j}\otimes R_{j,z},\\
D_j&=\sum_z\ket{z}\bra{z}_{\bar t_j}\otimes D_{j,z}.
\end{align}
Thus, $U=(D_{N-1}R_{N-1})\cdots(D_1R_1)$.

\subsection{Single-qubit synthesis}

We now describe how to implement a multiplexed single-qubit unitary using a Boolean function. For each control string $z$, the function gives a string of $H$ and $T$ instructions and a binary approximation of the remaining phase. We load this description, apply the instructions and phase correction, and then uncompute the description. Throughout, we assume access to ideal phase-gradient states.

\begin{lemma}[Programmed single-qubit synthesis]
\label{lem:singlesynth}
For a classically specified multiplexed single-qubit unitary
\begin{align*}
V=\sum_{z\in\{0,1\}^m}\ket{z}\bra{z}\otimes V_z
\end{align*}
and error bound $0<\delta<1/2$, there is a Boolean function $f:\{0,1\}^m\to\{0,1\}^L$, with $L=O(\log(1/\delta))$, describing the single-qubit operation for each branch. Given a Clifford+$T$ circuit $U_f$ implementing
\begin{align*}
\ket{z}\ket{0^L}\longmapsto\ket{z}\ket{f(z)},
\end{align*}
we can synthesize a $\delta$-approximation of $V$ in operator norm using $2T(U_f)+O(\log(1/\delta))$ $T$ gates and $O(\log(1/\delta))$ clean ancillas. 
\end{lemma}

\begin{proof}
First, write each branch as $V_z=e^{i\phi_z}R_z$, where $R_z\in SU(2)$. By single-qubit synthesis, we can approximate $R_z$ using a string $\widetilde R_z$ of $O(\log(1/\delta))$ $H$ and $T$ gates, up to a known phase $\gamma_z$ \cite{Ross16}:
\begin{align*}
\norm{\widetilde R_z-e^{i\gamma_z}R_z}\le\delta/2.
\end{align*}
We correct this extra phase when applying the remaining phase of $V_z$.

Let $\ell=O(\log(1/\delta))$ be a common bound on the string lengths. By inserting identity instructions, we can write each string as
\begin{align*}
\widetilde R_z=T^{b_{2\ell-1}(z)}H^{b_{2\ell-2}(z)}\cdots T^{b_1(z)}H^{b_0(z)},
\qquad b_i(z)\in\{0,1\}.
\end{align*}
Each bit of $b(z)$ tells us whether to apply its corresponding gate: $1$ means apply $H$ or $T$, and $0$ applies the identity. Thus, a bitstring $b(z)$ of length $2\ell$ programs the rotation.

We also need to apply the phase $\phi_z-\gamma_z$, which we approximate with bitstring $p_z$. Let $q=\lceil\log(2\pi/\delta)\rceil$ and choose a $q$-bit integer $p_z$ such that
\begin{align*}
\left|e^{2\pi i p_z/2^q}-e^{i(\phi_z-\gamma_z)}\right|\le\delta/2.
\end{align*}
We define $f(z)=b(z)p_z$, where $p_z$ is appended to the end of $b_z$, so its output has length $L=2\ell+q=O(\log(1/\delta))$.

The circuit proceeds as follows:
\begin{enumerate}
\item \textbf{Compute the description.} Apply $U_f$ to load $b(z)$ and $p_z$. This costs $T(U_f)$ $T$ gates.
\item \textbf{Apply the rotation.} For $i=0,\ldots,\ell-1$, apply $H$ controlled on $b_{2i}(z)$, followed by $T$ controlled on $b_{2i+1}(z)$. This step costs $O(\ell)$ $T$ gates.
\item \textbf{Apply the phase.} Add $p_z$ into the phase-gradient register
\begin{align*}
\ket{\psi_{\mathrm{PG}}}=2^{-q/2}\sum_{y=0}^{2^q-1}e^{-2\pi i y/2^q}\ket{y}
\end{align*}
to produce a phase-kickback:
\begin{align*}
\operatorname{ADD}_{p_z}\ket{\psi_{\mathrm{PG}}}=e^{2\pi i p_z/2^q}\ket{\psi_{\mathrm{PG}}}.
\end{align*}
The addition costs $O(q)$ $T$ gates \cite{Cuccaro04}.
\item \textbf{Uncompute the description.} Apply $U_f^\dagger$, costing another $T(U_f)$ $T$ gates. Since neither $z$ nor the description bits have changed, the description register returns to zero exactly.
\end{enumerate}

On branch $z$, we have implemented $\widetilde V_z=e^{2\pi i p_z/2^q}\widetilde R_z$. The rotation and phase errors are each at most $\delta/2$, so
\begin{align*}
\norm{\widetilde V_z-V_z}\le\delta.
\end{align*}
The error of the multiplexed unitary is the maximum branch error, rather than their sum, since it is block diagonal. Computing and uncomputing the boolean function costs $2T(U_f)$, and applying the rotation and phase costs $O(\ell+q)=O(\log(1/\delta))$ $T$ gates.
\end{proof}

\section{The Unary-Tracking Construction}\label{sec:main_sec}

We now prove Theorem~\ref{thm:main}, which we restate below for convenience.

\newtheorem*{restatedtheorem}{Theorem~\ref{thm:main}}
\begin{restatedtheorem}
Let $n\ge1$ and $0<\epsilon<1/2$. Given a classically specified $U\in U(N)$ with $K$ mixing branches in a supplied CSD factorization, there is a measurement-free Clifford+$T$ circuit $C$ using $a=N+O(L)$ clean ancillas and $T$-count
\begin{align*}
T(C)=O(K+NL).
\end{align*}
For every normalized input $\ket{\psi}$, it satisfies
\begin{align*}
\norm{C(\ket{\psi}\ket{0^a})-(U\ket{\psi})\ket{0^a}}\le\epsilon.
\end{align*}
We assume access to an ideal phase-gradient catalytic state.
\end{restatedtheorem}

\begin{proof}
We combine the unary lookup from Lemma~\ref{lem:lookup} with the programmed single-qubit synthesis from Lemma~\ref{lem:singlesynth}. We keep one $N$-qubit unary register $\mathsf A$ that tracks the data at the start of each CSD iteration. The lookups are free in terms of $T$-cost, and updating $\mathsf A$ throughout the CSD requires two Fredkin gates per mixing branch.

\subsection{Preparing the tracker}

First, compute the unary label of the input data into $\mathsf A$:
\begin{align}
\sum_x\alpha_x\ket{x}_{\mathsf D}\ket{0^N}_{\mathsf A}
&\longmapsto
\sum_x\alpha_x\ket{x}_{\mathsf D}\ket{e_x}_{\mathsf A}.
\label{eq:initial-tracker}
\end{align}
Start with an excitation on unary wire $0$ and read the data bits one at a time. At each level, Toffolis and CNOTs move the excitation to the wire for the corresponding binary prefix. There are $N-1$ splits, each costing $O(1)$ $T$ gates, so initialization costs $O(N)$ $T$ gates. We reverse this circuit at the end to clear the tracker.

\subsection{One CSD iteration}

Fix an iteration and write $t$ for its target qubit. For control string $z$, the unary wires $(A_{z,0},A_{z,1})$ correspond to target values $0$ and $1$. Using Lemma~\ref{lem:factor}, write the branch as $V_z=D_zR_z$. We implement it in the following steps.

\paragraph{1. Fold the tracker.}
For each mixing branch, apply a Fredkin gate to $(A_{z,0},A_{z,1})$, controlled on the data target $t$. If the target is $0$, the excitation is already on $A_{z,0}$. If it is $1$, the swap moves the excitation from $A_{z,1}$ to $A_{z,0}$. Thus,
\begin{align}
&\alpha_{z,0}\ket{z,0}_{\mathsf D}\ket{e_{z,0}}_{\mathsf A}
+\alpha_{z,1}\ket{z,1}_{\mathsf D}\ket{e_{z,1}}_{\mathsf A}
\nonumber\\
&\qquad\longmapsto
\bigl(\alpha_{z,0}\ket{z,0}+\alpha_{z,1}\ket{z,1}\bigr)_{\mathsf D}
\ket{e_{z,0}}_{\mathsf A}.
\label{eq:folded}
\end{align}
A Fredkin gate on an unoccupied pair does nothing. Diagonal and off-diagonal branches skip this step.

\paragraph{2. Load and apply the single-qubit operation.}
Using Lemma~\ref{lem:singlesynth}, choose the gate instructions and phase correction for each branch to accuracy $\delta$. On diagonal and off-diagonal branches, use exact target operations $I$ and $X$, respectively.

The gate instructions depend only on $z$, so we load them using CNOTs from the corresponding unary wire. The phase correction is also loaded from the tracker. For off-diagonal branches, we load the phase corresponding to the new target value. Next, we apply the programmed gates and phase-gradient addition, and then reverse the lookups. Since the tracker is unchanged throughout this step, the register containing the boolean function output gets cleaned up without garbage. The total cost is $O(\log(1/\delta))$ $T$ gates.

\paragraph{3. Restore the full unary label.}
We now update $\mathsf A$ to match the new data. On a diagonal branch, no update is needed. On an off-diagonal branch, apply an ordinary SWAP to $(A_{z,0},A_{z,1})$, since $X$ has flipped the target. On a mixing branch, apply a Fredkin gate to the same pair, controlled on the new target value.

Writing $\beta_{z,0}$ and $\beta_{z,1}$ for the amplitudes after the implemented rotation, the mixing-branch update gives
\begin{align}
&\bigl(\beta_{z,0}\ket{z,0}+\beta_{z,1}\ket{z,1}\bigr)_{\mathsf D}
\ket{e_{z,0}}_{\mathsf A}
\nonumber\\
&\qquad\longmapsto
\beta_{z,0}\ket{z,0}_{\mathsf D}\ket{e_{z,0}}_{\mathsf A}
+\beta_{z,1}\ket{z,1}_{\mathsf D}\ket{e_{z,1}}_{\mathsf A}.
\label{eq:restored}
\end{align}
The tracker therefore matches the data exactly for the rotation actually implemented, even when that rotation approximates the desired one.

At the end of the iteration, $\mathsf A$ tracks the updated data, and all intermediate registers are clean. We can therefore repeat the procedure for the next CSD factor.

\subsection{Completing the construction and counting resources}

After all $M=N-1$ factors, reverse the initial tracker computation to clear $\mathsf A$. All temporary workspace is cleared exactly, and the supplied phase-gradient state is unchanged.

Each mixing branch uses two Fredkin gates, each with $O(1)$ $T$-cost. The lookups and ordinary swaps use only Clifford gates. The programmed rotation and phase addition together cost $O(\log(1/\delta))$ $T$ gates per factor. For some iteration $j$, the $T$-cost is therefore $O(K_j+\log(1/\delta))$.

Choose $\delta=\epsilon/M$. By Lemma~\ref{lem:singlesynth}, each factor has error at most $\delta$, including its phase correction. The errors add across the $M$ factors, giving total error at most $\epsilon$.

Initializing and uncomputing the unary tracker costs $O(N)$ $T$ gates. Iterating through the whole CSD gives $T$-count:
\begin{align}
T(C)&=O\left(K+N+M\log\frac1\delta\right)\nonumber\\
&=O\left(K+2^n\left(n+\log\frac1\epsilon\right)\right).
\label{eq:main-bound}
\end{align}
The construction uses one $N$-qubit tracker and $O(\log(1/\delta))=O(L)$ clean temporary qubits, in addition to the supplied phase-gradient register. This gives $N+O(L)$ clean ancillas and $O(K+NL)$ $T$-count.
\end{proof}

\section{Space-Depth Volume Construction}
\label{sec:volume}

We now prove Theorem~\ref{thm:volume}. We transfer the quantum state into unary form so that different branches can run in parallel. To use fewer ancillas, we decompose the unitary into smaller blocks and implement each using the same reusable unary register.

\newtheorem*{restatedvolumetheorem}{Theorem~\ref{thm:volume}}
\begin{restatedvolumetheorem}
Let $N=2^n$ and $n\le a\le N$. Our construction uses at most $a$ ancillas to implement any $n$-qubit unitary exactly, using arbitrary single-qubit gates and CNOTs, and assuming unrestricted connectivity. Its depth and space-depth volume respectively are
\begin{align*}
D=O(N^2/a), \qquad V=(n+a)D=O(N^2).
\end{align*}
\end{restatedvolumetheorem}

\begin{proof}
We first describe the unary implementation of a small block, then use the quantum Shannon decomposition to reduce the workspace required. We count the depth of all elementary one- and two-qubit gates, with gates in the same layer acting on disjoint qubits.

\subsection{Implementing a block in unary form}

Consider a $k$-qubit block and let $B=2^k$. First, compute the unary label into a $B$-qubit register:
\begin{align*}
\sum_x\alpha_x\ket{x}_{\mathsf D}\ket{0^B}_{\mathsf A}
\longmapsto
\sum_x\alpha_x\ket{x}_{\mathsf D}\ket{e_x}_{\mathsf A}.
\end{align*}
We then clear the binary data using just CNOTs, so the state is now entirely held in the unary register:
\begin{align*}
\sum_x\alpha_x\ket{x}_{\mathsf D}\ket{e_x}_{\mathsf A}
\longmapsto
\ket{0^k}_{\mathsf D}\sum_x\alpha_x\ket{e_x}_{\mathsf A}.
\end{align*}
Computing the unary label takes $O(B)$ depth and clearing the binary data register also takes $O(B)$ depth.

We implement each branch $V_{j,z}$ directly on its pair of unary wires $(q_{z,0},q_{z,1})$. In this order, the required operation is
\begin{align*}
\ket{10}&\longmapsto (V_{j,z})_{00}\ket{10}+(V_{j,z})_{10}\ket{01},\\
\ket{01}&\longmapsto (V_{j,z})_{01}\ket{10}+(V_{j,z})_{11}\ket{01}.
\end{align*}
First, apply a CNOT from $q_{z,1}$ to $q_{z,0}$. This maps
\begin{align*}
\ket{10}\longmapsto\ket{10},\qquad
\ket{01}\longmapsto\ket{11}.
\end{align*}
On both states, $q_{z,0}$ is now $1$, while $q_{z,1}$ holds the target value. We can therefore apply $V_{j,z}$ to $q_{z,1}$, controlled on $q_{z,0}$. Reversing the CNOT returns the result to unary form.

For an unoccupied pair, the state is $\ket{00}$, so the controlled operation does nothing. Similarly, on input $\ket{11}$, the first CNOT flips the control bit to $0$ so no operation is performed. This procedure uses $O(1)$ gates and has depth $O(1)$.

Within a CSD factor, the pairs are disjoint and can run in parallel. There are $B-1$ factors, so implementing the block takes $O(B)$ depth. Including encoding and decoding, one block therefore has depth $O(B)$. We maintain the unary register throughout so it can be uncomputed back into the data register at the end of the procedure.

\subsection{Reducing the workspace using QSD}

The quantum Shannon decomposition (QSD) writes a $j$-qubit unitary as
\begin{align}
U=(I\otimes W_1)M_z^{(1)}(I\otimes W_2)M_y(I\otimes W_3)M_z^{(2)}(I\otimes W_4),
\label{eq:qsd}
\end{align}
where each $W_i$ is an uncontrolled $(j-1)$-qubit unitary and $M_z$ and $M_y$ are multiplexors with the same $j-1$ qubits as the control string and the target is the remaining qubit \cite{SMB06}. We recursively decompose each $W_i$, stopping when the smaller blocks fit within our qubit budget.

Given an ancilla budget $a$, choose
\begin{align*}
k=\lfloor\log a\rfloor,\qquad B=2^k,
\end{align*}
and use a $B$-qubit unary register, which is $\leq a$. 

We stop the QSD recursion at $k$-qubit blocks. Each level produces four smaller blocks, so after $n-k$ levels there are
\begin{align}
4^{n-k}=(N/B)^2
\end{align}
blocks. For each block, we encode its $k$ data qubits into a $B$-qubit unary, implement the block following the procedure from the previous subsection, and decode back into binary form. Decoding clears the unary register, allowing us to reuse the same $B$ qubits for the next block.

Each block, including encoding and decoding, takes $O(B)$ depth. Executing all blocks sequentially therefore contributes
\begin{align*}
O\!\left((N/B)^2B\right)=O(N^2/B)
\end{align*}
depth. We count the multiplexed rotations between the blocks separately below.

\subsection{Counting depth and volume}

We also include the multiplexed rotations between the blocks. At a $j$-qubit QSD node, the three rotations can be implemented on the binary data using $O(2^j)$ depth and no additional ancillas \cite{SMB06}. There are $4^{n-j}$ such nodes, giving total rotation depth
\begin{align*}
O\!\left(\sum_{j=k+1}^{n}4^{n-j}2^j\right)=O(N^2/B).
\end{align*}
Combining this with the block implementations, including every encoding and decoding, gives
\begin{align*}
D\in O(N^2/B) \subseteq O(N^2/a).
\end{align*}

The circuit uses $n$ data qubits and at most $a$ ancillas. Since $a\ge n$, its space-depth volume is
\begin{align*}
V=(n+a)D
=O\!\left((n+a)\frac{N^2}{a}\right)
=O(N^2).
\end{align*}
The unary register is cleared after every block, and the intervening rotations require no additional workspace. All ancillas therefore return to zero exactly.
\end{proof}

This volume is asymptotically optimal in the worst case. A generic $n$-qubit unitary requires $\Omega(N^2)$ continuous parameters, while each elementary gate supplies only $O(1)$ parameters \cite{SMB06}. Since a layer contains at most $O(n+a)$ gates, the worst-case space-depth volume is $\Omega(N^2)$.

\section{Conclusion}\label{sec:conclusion}

We presented a unitary synthesis construction using a unary register to track the computational-basis label of the data. This makes Boolean lookups free in $T$-count and lets us share the precision-dependent synthesis cost across all branches of a CSD factor. Assuming access to ideal phase-gradient states, the construction uses $N+O(L)$ clean ancillas and $O(K+NL)$ $T$ gates, where $K$ counts the mixing branches in the CSD of the unitary we are synthesizing.

We also described a variant that directly applies the unitary onto a unary register and applies the disjoint branches in parallel. Using QSD to divide the unitary into smaller blocks lets us reuse a smaller unary register. With arbitrary single-qubit gates and CNOTs, this gives depth $O(N^2/a)$ and space-depth volume $O(N^2)$ for $n\le a\le N$, matching the known asymptotically optimal volume bound.

The usefulness of the mixing-sensitive bound depends on finding factorizations with small $K$. An interesting direction is to identify classes of unitaries that admit such factorizations, or to develop methods for reducing $K$. It would also be useful to characterize this construction in terms of its concrete fault-tolerant resource costs for practical applications.

For the $T$-count construction, the main open question is whether the unary tracker can be updated more cheaply. Our procedure uses $O(K_j)$ $T$ gates in factor $j$. Can we reduce this to a cost polynomial in $n$, while preserving the zero-$T$ Boolean lookups and exact cleanup? This may require a different encoding or a way to update many unary pairs together. Solving this open question would achieve an a $\widetilde{O}(2^n)$ $T$-count bound for general unitary synthesis. 

A related question is whether there exists a more efficient fault-tolerant implementation of the updating to the unary tracker. In the worst case, each factor of the CSD requires $2^{n-1}$ Fredkin gates to update. However, these Fredkins are all controlled on the same qubit, but act on disjoint pairs. An interesting direction to explore is whether this \textit{Wide Fredkin} admits a more efficient decomposition than directly implementing each gate independently. Exploiting this structure could reduce the $T$-count or the cost of the required magic-state factories, even without changing the tracker.

\begin{acknowledgments}
The author thanks Richard Cleve, Gregory Rosenthal, Luke Schaeffer, Kewen Wu, Guillermo Alonso-Linaje, and David Wierichs for the many insightful discussions that helped motivate this project. This work was supported in part by an Undergraduate Research Assistantship (URA) from the University of Waterloo.

\end{acknowledgments}

\section*{Use of Generative AI}

The author used OpenAI's ChatGPT to assist with editing the manuscript and checking mathematical arguments. All ideas and constructions were independently developed by the author. The author takes full responsibility for the final results and conclusions.

\bibliography{unitary_synthesis}

@misc{Tan25,
    title={{Unitary synthesis with fewer T gates}},
    author={Xinyu Tan},
    year={2025},
    eprint={2509.25702},
    archivePrefix={arXiv},
    primaryClass={quant-ph},
    url={https://arxiv.org/abs/2509.25702},
}

@article{GKW24,
    title={{Quantum state preparation with optimal T-count}},
    volume={10},
    ISSN={2521-327X},
    url={http://dx.doi.org/10.22331/q-2026-07-22-2168},
    DOI={10.22331/q-2026-07-22-2168},
    journal={Quantum},
    publisher={Verein zur Forderung des Open Access Publizierens in den Quantenwissenschaften},
    author={Gosset, David and Kothari, Robin and Wu, Kewen},
    year={2026},
    month=July,
    pages={2168},
}

@misc{Ross16,
    title={{Optimal ancilla-free Clifford+T approximation of z-rotations}},
    author={Neil J. Ross and Peter Selinger},
    year={2016},
    eprint={1403.2975},
    archivePrefix={arXiv},
    primaryClass={quant-ph},
    url={https://arxiv.org/abs/1403.2975},
}

@article{Litinski19,
    title={{A Game of Surface Codes: Large-Scale Quantum Computing with Lattice Surgery}},
    volume={3},
    ISSN={2521-327X},
    url={http://dx.doi.org/10.22331/q-2019-03-05-128},
    DOI={10.22331/q-2019-03-05-128},
    journal={Quantum},
    publisher={Verein zur Forderung des Open Access Publizierens in den Quantenwissenschaften},
    author={Litinski, Daniel},
    year={2019},
    month=Mar,
    pages={128},
}

@article{LKS24,
    title={{Trading T gates for dirty qubits in state preparation and unitary synthesis}},
    volume={8},
    ISSN={2521-327X},
    url={http://dx.doi.org/10.22331/q-2024-06-17-1375},
    DOI={10.22331/q-2024-06-17-1375},
    journal={Quantum},
    publisher={Verein zur Forderung des Open Access Publizierens in den Quantenwissenschaften},
    author={Low, Guang Hao and Kliuchnikov, Vadym and Schaeffer, Luke},
    year={2024},
    month=June,
    pages={1375},
}

@misc{YZZ26,
    title={{Quantum Circuit for General Unitary: Improved T-count via Block Flattening and Dilation}},
    author={Pei Yuan and Shengyu Zhang and Wei Zi},
    year={2026},
    eprint={2608.17846},
    archivePrefix={arXiv},
    primaryClass={quant-ph},
    url={https://arxiv.org/abs/2608.17846},
}

@misc{Liu26,
    title={{$T$-efficient unitary synthesis via block-sparse Hamiltonian simulation}},
    author={Diyi Liu and Shuchen Zhu and Wibe Albert de Jong and Di Fang and David Williams-Young and Chao Yang},
    year={2026},
    eprint={2610.18589},
    archivePrefix={arXiv},
    primaryClass={quant-ph},
    url={https://arxiv.org/abs/2610.18589},
}

@article{SMB06,
    title={{Synthesis of quantum-logic circuits}},
    volume={25},
    ISSN={1937-4151},
    url={http://dx.doi.org/10.1109/TCAD.2005.855930},
    DOI={10.1109/tcad.2005.855930},
    number={6},
    journal={IEEE Transactions on Computer-Aided Design of Integrated Circuits and Systems},
    publisher={Institute of Electrical and Electronics Engineers (IEEE)},
    author={Shende, V.V. and Bullock, S.S. and Markov, I.L.},
    year={2006},
    month=June,
    pages={1000–1010},
}

@misc{Sun23,
    title={{Asymptotically Optimal Circuit Depth for Quantum State Preparation and General Unitary Synthesis}},
    author={Xiaoming Sun and Guojing Tian and Shuai Yang and Pei Yuan and Shengyu Zhang},
    year={2023},
    eprint={2108.06150},
    archivePrefix={arXiv},
    primaryClass={quant-ph},
    url={https://arxiv.org/abs/2108.06150},
}

@misc{Yuan26,
    title={{Exact Unitary Synthesis: Simultaneously Optimal in Size and Depth-Ancilla Tradeoff}},
    author={Pei Yuan and Wei Zi and Shengyu Zhang},
    year={2026},
    eprint={2610.07010},
    archivePrefix={arXiv},
    primaryClass={quant-ph},
    url={https://arxiv.org/abs/2610.07010},
}

@misc{Cuccaro04,
    title={{A new quantum ripple-carry addition circuit}},
    author={Steven A. Cuccaro and Thomas G. Draper and Samuel A. Kutin and David Petrie Moulton},
    year={2004},
    eprint={quant-ph/0410184},
    archivePrefix={arXiv},
    primaryClass={quant-ph},
    url={https://arxiv.org/abs/quant-ph/0410184},
}

\end{document}